\documentclass[aps,preprint,nofootinbib]{revtex4}%
\usepackage{amsfonts}
\usepackage{amsmath}
\usepackage{amssymb}
\usepackage{float}
\usepackage{romannum}
\usepackage{graphicx}%
\usepackage{color}
\usepackage{enumitem}

\providecommand{\U}[1]{\protect\rule{.1in}{.1in}}

\begin{document}


\title{Endorsing black holes with beyond Horndeski primary hair: An exact solution framework for scalarizing in every dimension}


\author{Olaf Baake$^{1}$, Adolfo Cisterna$^{2}$, Mokhtar Hassaine$^{1}$ and Ulises Hernandez-Vera$^{1}$}

\affiliation{$^{1}$Instituto de Matem\'{a}tica y F\'{\i}sica,
Universidad de Talca, Casilla 747,Talca, Chile}

\affiliation{$^{2}$Institute of Theoretical Physics, Faculty of
Mathematics and Physics, Charles University, V Holesovickach 2, 180
00 Prague 8, Czech Republic}

\email{olaf.baake@gmail.com, adolfo.cisterna@mff.cuni.cz,
hassaine@inst-mat.utalca.cl, uhernandez.vera@gmail.com}

\begin{abstract}
This work outlines a straightforward mechanism for endorsing primary
hair into Schwarzschild black holes, resulting in a unique
modification within the framework of a special scalar-tensor theory, the so-called beyond Horndeski type. The derived solutions are
exact, showcase primary hair with an everywhere regular scalar field
profile, and continuously connect with the vacuum geometry. Initially
devised to introduce primary hair in spherically symmetric solutions
within General Relativity in any dimension, our
investigation also explores the conditions under which spherically
symmetric black holes in alternative gravitational theories become
amenable to the endorsement of primary hair through a similar
pattern. As a preliminary exploration, we embark on the process of
endorsing primary hair to the Reissner-Nordström black hole. Subsequently, we extend
our analysis to encompass spherically symmetric solutions within
Lovelock and cubic quasitopological gravity theories.
\noindent
\end{abstract}

\maketitle

\section{Introduction}

Undoubtedly, the present era constitutes a remarkable epoch for the
exploration of gravitational phenomena. The affirmation of
gravitational waves \cite{GW}, the first black hole image
achieved by the Event Horizon Telescope network \cite{EHT}, and the
investigation into the precession orbits of stars revolving around
compact massive objects \cite{stars}, collectively present an
unprecedented opportunity to scrutinize gravity on a stage far
exceeding the well-established scales of the Solar System
\cite{Will:2014kxa}. Consequently, a discernible dichotomy emerges:
Einstein's theory of General Relativity (GR) assumes an unequivocal
primacy, having demonstrated confirmation in the realm of strong
gravity. Simultaneously, a newfound avenue emerges for the
evaluation of alternative gravity theories aspiring to enhance GR
and provide solutions to phenomena where the theory proves
inadequate. In this context, theories incorporating additional
degrees of freedom, particularly scalar fields, emerge as economical
and archetypal modifications of GR. Since their inception, these
models, encapsulated within the framework of Scalar-Tensor theories
of gravity, have undergone extensive scrutiny, primarily in the
realm of Cosmology \cite{Clifton:2011jh}, though their applicability
extends far beyond. These models find their most comprehensive
formulation in Horndeski theory \cite{Horndeski:1974wa}, alongside
subsequent higher-order modifications such as Beyond Horndeski
theories \cite{Gleyzes:2014dya,Gleyzes:2014qga} and Degenerated
Higher Order Scalar Tensor theories (DHOST)
\cite{Langlois:2015cwa,Langlois:2015skt,Crisostomi:2016czh,BenAchour:2016fzp}.
Establishing a robust foundation for the validity of these
alternative models necessitates a meticulous examination of their
spectrum of black hole solutions. This endeavor not only furnishes a
theoretical framework for assessing the consistency of these models
but also anticipates potential experimental implications.

It is commonly argued that, following gravitational collapse, a
black hole can be adequately described by a specific set of
parameters, namely its mass, electromagnetic charges, and angular
momentum. This perspective implies that no additional distinctive
features of the original matter persist after the black hole
formation, such as baryon or lepton numbers. In certain
scenarios, this proposition is substantiated, leading to the
formulation of no-hair theorems \cite{Ruffini:1971bza}. Here, the
term "hair" is used metaphorically to encompass all characteristics
that would render black holes non-bald, indicating quantities not
subject to a Gauss law and, consequently, not conserved at infinity.
This definition originates from the numerical construction of
Einstein-Yang-Mills black holes, where a discrete parameter, that
represent the number of nodes of the gauge function \cite{EYM,
Bizon:1994dh}, emerges.

Over the decades, numerous endeavors have been undertaken to
construct solutions introducing various types of hairs to black hole
configurations. Notably, there has been a particular focus on scalar
hair, which involves configurations where a nontrivial scalar field
profile coexists within the spacetime, primarily driven by the
assumption of the inherent simplicity of scalar fields
\cite{Herdeiro:2015waa}. However, the incorporation of scalar fields
into black hole configurations proves to be challenging, with many
solutions exhibiting curvature singularities or divergence of the
scalar field profile within the domain of outer communications. As
non-hair theorems constitute theory-dependent mathematical
statements, the scientific community has continuously subjected them
to scrutiny to assess their validity. Early contributions by Chase
\cite{Chase} and subsequently by Bekenstein \cite{Bekenstein:1972ny}
defined conditions under which minimally coupled scalar fields
cannot dress a black hole spacetime. Similar no-hair theorems were
also formulated by Hawking for the Brans-Dicke theory
\cite{Hawking:1972qk}, later generalized in the presence of
self-interaction for the scalar field \cite{Sotiriou:2011dz}. In a
more contemporary context, modern scalar-tensor theories,
predominantly represented by Horndeski gravity and its higher-order
extensions beyond Horndeski and DHOST theories, have faced
constraints concerning the emergence of hairy black hole solutions
\cite{Hui:2012qt,Maselli:2015yva,Babichev:2016rlq,Creminelli:2020lxn}.
However, the potency of these theorems is contingent upon their
underlying assumptions, and it is, therefore, plausible to
circumvent them by precisely relaxing some of their key axioms. A
substantial body of literature has emerged, delving into the
construction and investigation of black holes with scalar hair. This
journey began with the discovery of black holes in the context of
conformally coupled scalar theories \cite{Bekenstein:1974sf,BBM,
Bekenstein:1975ts,Xanthopoulos:1992fm,Martinez:2002ru,Martinez:2005di,
Charmousis:2009cm,Bardoux:2013swa,Anabalon:2009qt,Barrientos:2023tqb,
Cisterna:2021xxq,Caceres:2020myr,Anabalon:2012tu,Barcelo:2000zf,Ayon-Beato:2015ada,Barrientos:2016ubi}
and has progressed to encompass more recent configurations found in
complex minimally coupled models \cite{Herdeiro:2014goa}, and
Horndeski as well as higher-order scalar-tensor theories
\cite{Rinaldi:2012vy,Babichev:2013cya,Anabalon:2013oea,Cisterna:2014nua,
Minamitsuji:2013ura,Kobayashi:2014eva,Minamitsuji:2018vuw,Motohashi:2019sen,
Minamitsuji:2019shy,Minamitsuji:2019tet,BenAchour:2018dap,BenAchour:2020wiw,
BenAchour:2020fgy,Charmousis:2019vnf,Anson:2020trg,Baake:2021jzv,Bakopoulos:2022csr,
Babichev:2022awg,Babichev:2023dhs,Babichev:2023psy,Chatzifotis:2021hpg,Bakopoulos:2023tso,Babichev:2017guv,Bakopoulos:2021liw,Charmousis:2021npl,Babichev:2020qpr}.

The diversification of hairy black hole solutions has led to a
nuanced understanding of the scalar hair concept, culminating in the
categorization of two distinct types: primary and secondary. The
latter denotes black hole spacetimes characterized by a non-trivial
scalar field profile, which, crucially, does not introduce any
additional parameter to the geometry. Consequently, the backreaction
of the spacetime exhibits no explicit manifestation of the scalar
hair, preventing these solutions from forming a continuous
connection with vacuum black hole spacetimes. In contrast, primary
hair designates black hole spacetimes featuring a nontrivial scalar
field configuration that alters the spacetime backreaction by
incorporating an additional parameter. Consequently, black holes
with primary hair can form a continuous connection with vacuum
solutions, that is, with their bald counterpart geometries.
While solutions with secondary hair constitute the majority of
existing exact solutions documented in the literature, those with
primary hair have predominantly been constructed through numerical
methods \cite{Herdeiro:2014goa,Herdeiro:2016tmi}.

A recent proposal has introduced an intriguing mechanism for the
systematic numerical construction of black holes characterized by
primary hair. This process known as "scalarization" represents a
pathway by which a vacuum black hole can develop scalar hair through
a tachyonic instability, revealing the emergence of black holes with
primary hair at its culmination. The fundamental properties defining
a black hole with primary hair include the presence of black holes
exhibiting a consistently regular scalar field configuration.
Additionally, their backreaction is contingent upon the existence of
a scalar charge (a continuous parameter governing the manifestation
of the scalar field profile). These black hole solutions with
primary hair have undergone thorough investigation in recent years,
encompassing not only spherically symmetric configurations
\cite{Doneva:2017bvd,Doneva:2018rou,Silva:2017uqg,Guo:2020sdu,Kiorpelidi:2023jjw,Minamitsuji:2018xde}
but also extending to stationary and axially symmetric ones
\cite{Brihaye:2018bgc,Herdeiro:2020wei,Berti:2020kgk}. Furthermore, this
exploration has extended to hairs beyond the scalar variety,
including vectorial and tensorial natures
\cite{Brihaye:2020oxh,Brihaye:2019kvj,Barton:2021wfj,Ramazanoglu:2017xbl,Minamitsuji:2020pak,Ikeda:2019okp,Ramazanoglu:2019gbz}.

Efforts directed towards the construction of exact black hole
solutions featuring primary hair remain limited. This scarcity
primarily stems from the intricate nature of the theories within
which the search for such hair is conducted, compounded by the
challenges posed by the underlying complexities embedded in no-hair
theorems. Furthermore, scalarizing processes in the presence of a
cosmological constant or within an arbitrary number of dimensions
pose considerable difficulties and represent a direction far less
explored \cite{Brihaye:2019dck,Brihaye:2019gla}. These constraints
significantly limit the applicability of these black holes for
exploration, particularly in realms such as black hole
thermodynamics or other semiclassical phenomena within the framework
of the AdS/CFT conjecture
\cite{termo1,Maldacena,HPT,Witten:1998qj,witten2} just to name a few
examples. Consequently, the pursuit of exact black hole solutions
with primary hair, or, due to the similarities in
the process, exact scalarized black holes\footnote{The determination
of whether these exact black hole solutions with primary hair
exhibit a tachyonic instability is not the primary focus of this
comparison. Instead, we emphasize the significance of a black hole
possessing primary hair as the key characteristic for it to be
regarded as scalarized.}, becomes an intriguing avenue of
investigation.

Recently, analytical black hole solutions featuring primary scalar
hair were discovered in \cite{Bakopoulos:2023fmv} within the
framework of beyond Horndeski theories \cite{Gleyzes:2014dya} in
four dimensions. Beyond Horndeski theories represent extensions of
the well-known Horndeski theories \cite{Horndeski:1974wa},
incorporating higher-order derivatives while avoiding Ostrogradski
ghosts. These theories have demonstrated considerable promise in the
exploration of compact objects, as evidenced in works such as
\cite{Babichev:2017guv, Bakopoulos:2021liw, Charmousis:2021npl}.
This promise extends to the construction of scalarized black holes
or black holes with primary hair. In particular, the authors of
\cite{Bakopoulos:2023fmv} demonstrated the existence of an extension
of the Schwarzschild black hole within precise beyond Horndeski
models, specifically when $G_2\sim X^2$, $G_4\sim X^2$, and $F_4\sim
cte$ (see the action (\ref{actionbH}) below). Remarkably, the
resulting spacetime remains described by the Schwarzschild metric,
augmented by a term proportional to the scalar hair. Furthermore, it
was revealed that the inclusion of this additional term enables the
elimination of the central singularity through a specific tuning
between the mass and the hair.

In this study, we aim to broaden the findings established in
\cite{Bakopoulos:2023fmv} by extending the class of beyond Horndeski
theories capable of accommodating similar black hole solutions
endowed with primary scalar hair. Despite the original theory being
defined in four dimensions, our investigation will encompass the
arbitrary dimensional case. Our focus is on demonstrating that the
existence of hairy solutions, extending the Schwarzschild black hole
paradigm, can be guaranteed for a broader selection of theory
functions. Specifically, we will establish that a two-parametric
subclass of actions, characterized by functions $G_2$ and $G_4$
within the framework of (\ref{actionbH}), facilitates the emergence
of such scalarized solutions. The resulting metric solution comprises a
superposition of the Schwarzschild-(A)dS function with an additional
component proportional to the scalar hair. We will further expand on
this pattern by demonstrating that, under specific hypotheses, this
two-parametric class of beyond Horndeski theories can be coupled
with other actions of pure gravity (beyond GR and potentially
involving additional dynamical fields). This coupling allows for the
extension of purely static black hole solutions to static black
holes with primary hair.

This paper is organized as follows: In Section II, we elaborate our
approach to endorsing vacuum black holes with primary scalar hair.
In essence, we outline the generic construction of scalarized black
holes within the domain of spherically symmetric solutions. We delve
into the explicit construction of exact scalarized Schwarzschild
black holes within a theory characterized by the form
(\ref{actionbH}). Specifically, we demonstrate that solutions
featuring primary hair and smoothly connecting to the
Schwarzschild-(A)dS solution may exist for a subset of actions
(\ref{actionbH}) parameterized by $G_2$ and $G_4$. The specific
scenario where both coupling functions are proportional is examined
in detail. Section III introduces a set of conditions under which
the framework outlined in Section II can be extended to encompass
other gravity theories, whether purely geometrical or involving
additional matter fields (distinct from the beyond Horndeski scalar field, $\phi$). We illustrate how black holes in alternative theories can
be enhanced to exhibit primary hair of the beyond Horndeski type, as
defined by (\ref{actionbH}). To illustrate this, we demonstrate how the
Einstein-Maxwell theory supports black holes with primary hair,
thereby explicitly constructing a scalarized version of the
Reissner-Nordström black hole. Subsequently, we extend our
exploration to the construction of black holes with primary hair in
Lovelock and cubic quasitopological gravities. Finally, Section IV
is dedicated to concluding and proposing several avenues for further
exploration and generalization of the framework presented herein.
Given the generic nature of our approach, we allocate an Appendix dedicated to the construction of specific black hole configurations. In particular, we delve into the cases of GR and the Einstein-Gauss-Bonnet theory.

\section{Scalarizing the Schwarzschild black hole}
The primary objective of this work is to introduce primary hair onto
initially bald black hole solutions, beginning within the framework
of General Relativity (GR) and subsequently extending to other
geometric theories of gravity. Despite the extensive literature on
the construction of black holes with hair, these solutions typically
manifest secondary hair, exemplified by stealth black holes or
standard black holes lacking a continuous limit with the vacuum
(bald) geometry. Various techniques have been employed to construct
these solutions, ranging from scalar fields that do not share the
same symmetries as the geometry to the utilization of disformal
transformations. A prevalent characteristic of most solutions is
their emergence within theories featuring shift symmetry. This
allows the scalar field equation to be formulated as a current
conservation law, facilitating the integration of field equations.
Another commonly adopted strategy involves stipulating a constant
kinetic term for the scalar field. This considerably simplifies the
contribution of the scalar sector within a given scalar-tensor
theory, reducing the problem to finding stealth black holes by
adjusting the Lagrangian functions. Such solutions are
attainable when the scalar field profile exhibits a linear
time-dependence, a characteristic that, due to the shift
invariant nature of the models, does not compromise the stationary
nature of the solutions. More challenging is the discovery of
solutions with a non-constant kinetic term, representing black holes
with novel backreactions. In the subsequent sections, we will
assimilate various elements from the existing literature, and combine
them in a manner that facilitates the systematic construction of
exact black hole solutions with primary hair. This approach is
applicable to a sufficiently general class of theories encapsulated
within beyond Horndeski gravity and will generically contain a
non-trivial kinetic term for the scalar field.

\subsection{Beyond Horndeski theory and the scalarization scheme}
As our initial aim is to scalarize the Schwarzschild black hole and
its higher dimensional extension, the Schwarzschild-Tangherlini
black hole, and thus we consider the action of quadratic beyond Horndeski
gravity \cite{Gleyzes:2014dya} elevated to an arbitrary dimension, $d$:
\begin{align}\label{actionbH}
\begin{split}
S&=\int
d^dx\sqrt{-g}\Bigg[G_2(X)+G_4(X)R+\left(G_{4,X} + 2\,X F_4(X)\right)\left(\left(\Box\phi\right)^2
-\phi_{\mu\nu}\phi^{\mu\nu}\right) \\
&\qquad\qquad\qquad\qquad\qquad + 2 F_4(X)\Big(\square \phi \phi^\mu \phi_{\mu \nu} \phi^\nu -  \phi_\mu \phi^{\mu \nu} 
\phi_{\nu \rho} \phi^\rho\Big)\Bigg].
\end{split}
\end{align}
For simplicity, we have defined $\phi_{\mu}=\partial_{\mu}\phi$ and
$\phi_{\mu\nu}=\nabla_{\mu}\nabla_{\nu}\phi$, where the coupling
functions $G_2$, $G_4$ and $F_4$ depend solely on the kinetic term
$X=-\frac{1}{2}\phi_{\mu}\phi^{\mu}$. Here $G_{4,X}$ stands for the
derivative of $G_4$ with respect to $X$, i.e. $G_{4,X}=\frac{d
G_4(X)}{dX}$, and $\epsilon_{\mu\nu\rho\sigma}$ stands for the
Levi-Civita tensor. Notice that GR is naturally included by a proper
choice of the theory function $G_4$. Action (\ref{actionbH}) is
invariant under a constant shift of the scalar field
$\phi\to\phi+\mbox{cst}$, a heritage from the Galileon origin of the
model \cite{Nicolis:2008in}, and it is parity invariant $\phi\to
-\phi$ as it is quadratic in the derivatives of the scalar field.

Focusing on spherical symmetry, we consider a $d$-dimensional
spacetime configuration of the form
\begin{equation}
ds^2=-h(r)dt^2+\frac{dr^2}{f(r)}+r^2d\Omega_{d-2,\kappa}^2\hspace{0.1cm},\quad\phi(t,r)=qt+\psi(r),
\label{ansatz}
\end{equation}
where the $(d-2)$-dimensional base manifold has constant curvature
$\kappa=0,\pm 1$ representing a spherical, hyperbolic or flat
topology, respectively. Here, $q$ is a constant of integration that
will also appear in the black hole metric function
\cite{Bakopoulos:2023fmv}, which will give the constant $q$ the
character of primary hair. In addition, it plays a crucial role in
the regularity of the scalar field profile. As already mentioned,
the action (\ref{actionbH}) enjoys invariance under a constant
translation of the scalar field, and as a consequence, the scalar
field equation of motion converts into a conservation law for the
scalar Noether current,
$$
\mathcal{J}^{\mu}=\frac{1}{\sqrt{-g}}\frac{\delta
S}{\delta(\partial_{\mu}\phi)},\qquad
\qquad\nabla_{\mu}\mathcal{J}^{\mu}=0.
$$
This property has played a major role in the construction of black
hole solutions in Horndeski gravity and its higher order
generalizations. Its regularity at the would-be black hole horizon
along with a few other assumptions regarding the asymptotic behavior
of the would-be solutions and the analyticity of the theory's
Lagrangian constitute the cornerstone of no hair theorems, and
therefore almost by transitivity has paved the road to understand
how these theory dependent statements can be circumvented to obtain
interesting geometries featuring non-trivial hair.

Along the lines of \cite{Babichev:2015rva}, it is possible to show
that for a configuration of the form (\ref{ansatz}) the independent
field equations to solve reduce to the metric variation equations
$\epsilon_{tt}=0$ and $\epsilon_{rr}=0$ and the vanishing of the
radial Noether current $\mathcal{J}^{r}=0$. As a matter of fact, the
non-diagonal Einstein equation $\epsilon_{tr}=0$ sourced by the
linear time dependence of the scalar profile, turns out to be
proportional to the scalar current $\mathcal{J}^{r}$ and thus no
flux for the scalar field takes place. As a consequence, the field
equations of our theory take the convenient form
\begin{align}
\mathcal{J}^r&:=r^2 h^2 \tilde{G}_{2, X}+(d-2)(d-3)\left(\kappa h^2-\frac{q^2 f h}{2 X}\right) \tilde{G}_{4 X}+(d-2) q^2 h^2\left(\frac{f}{h}\right)^{\prime} r F_4\label{eqJ} \\
&-(d-2)\left((d-3) f h^2+f h h^{\prime} r-\frac{(d-3) q^2 f h}{2 X}\right) \mathcal{Z}_X  \nonumber\\
\begin{split} \label{eqrr}
\epsilon_{r r}&:=h^3\left[-\frac{(d-2) f h^{\prime}}{h} r \mathcal{Z}-r^2\left(a_0+\tilde{G}_2\right)-(d-2)(d-3) \kappa\left(a_1+\tilde{G}_4\right)\right. \\
&\left.-(d-2)(d-3) f \mathcal{Z}+\frac{(d-2)(d-3) q^2 f}{2 X h}\left(\mathcal{Z}+a_1+\tilde{G}_4\right)-\frac{2(d-2) q^2 f}{h} r F_4 X^{\prime}\right]\\
&-\left(q^2-2 h X\right) \mathcal{J}^r
\end{split} \\
\epsilon_{t t}&:=-\epsilon_{r r}-2\left(q^2-h X\right)
\mathcal{J}^r+2 r^2 X^{\prime}
\mathcal{Z}_X\left(\frac{h}{f}\right)-r^2
\mathcal{Z}\left(\frac{h}{f}\right)^{\prime}. \label{eqtt}
\end{align}
Please note that for convenience, we have rescaled the equations in
the following way:
\begin{align*}
\epsilon_{tt} \rightarrow 2 r^2 \epsilon_{tt},\qquad \epsilon_{rr}
\rightarrow 2 r^2 h^3 \epsilon_{rr},\qquad
    \mathcal{J}^r &\rightarrow - \frac{r^2 h^3}{f
    \psi'}\mathcal{J}^r.
\end{align*}
Further, we have voluntarily re-written the coupling functions $G_2$
and $G_4$ as
\begin{equation}
G_2(X)=a_0+\tilde{G}_2(X), \quad G_4(X)=a_1+\tilde{G}_4(X), \quad
\tilde{G}_{4, X} \neq 0,  \label{conditions}
\end{equation}
and hence the constant $a_0$ represents an eventual bare
cosmological constant, while $a_1$ corresponds to the
standard Einstein-Hilbert term in the action. In addition, it turns
out to be advantageous to define the auxiliary function
\begin{equation}
\mathcal{Z}(X):=4 X^2 F_4+2 X \tilde{G}_{4,
X}-\left(a_1+\tilde{G}_4\right).
\end{equation}

From these independent equations, one can easily visualize the
emergence of hairy (scalarized) extensions of the Schwarzschild
black hole. Indeed, considering the homogeneous static case $f=h$ in
(\ref{ansatz}), the compatibility of the last equation (\ref{eqtt})
guides us towards two options, imposing either $\mathcal{Z}$ or $X$
to be constant, and here, we will consider  the first possibility,
$\mathcal{Z}=$ cst. The case of constant $X$, as it is known,
naturally leads to the construction of stealth black hole solutions.
Hence, for $\mathcal{Z}=$ cst $=\mathcal{Z}_0$, the radial current
equation $\mathcal{J}^r=0$ given by (\ref{eqJ}) reduces to the
simple expression
\begin{equation}
r^2 \tilde{G}_{2, X}+(d-2)(d-3)\left(\kappa-\frac{q^2}{2 X}\right)
\tilde{G}_{4, X}=0, \label{Jmaster}
\end{equation}
which later will provide the specific radial dependence (at least
implicitly) of the kinetic term $X=X(r)$. Next, choosing the
constant function $\mathcal{Z}$ to be $\mathcal{Z}=-a_1$, the
remaining independent equation $\epsilon_{r r}=0$ factorizes in the
very suitable form
\begin{align}
\begin{split}\label{master}
&-a_0 r^2+a_1(d-2)\left[r h^{\prime}+(d-3) h-\kappa(d-3)\right]-2(d-2) q^2 r F_4 X^{\prime}-r^2 \tilde{G}_2 \\
&\qquad\qquad\qquad\qquad\qquad\qquad\qquad\qquad-(d-2)(d-3)\left(\kappa-\frac{q^2}{2
X}\right) \tilde{G}_4=0 .
\end{split}
\end{align}
From here the following observations are in order: i) the terms
proportional to $a_0$ and $a_1$ will vanish identically for a
Schwarzschild (A)dS metric function $h$, (ii) the term involving
$X^{\prime}$ represents a sort of non-homogeneity, and (iii) the
last two terms of the equation (\ref{master}) are a "kind" of first
integral with respect to $X$ of the equation (\ref{Jmaster}). In
fact, using equation (\ref{Jmaster}), equation (\ref{master}) can be written
as
\begin{align}
\begin{split}\label{comerr2}
-a_0 r^2+a_1(d-2)\Bigg[rh'+(d-3)h-\kappa (d-3)\Bigg]-2 (d-2) q^2  r
F_4
X'\\+\tilde{G}_{4,X}(d-2)(d-3)\left(\kappa-\frac{q^2}{2X}\right)\Big(\frac{\tilde{G}_2}{\tilde{G}_{2,X}}-\frac{\tilde{G}_4}{\tilde{G}_{4,X}}\Big)=0,
\end{split}
\end{align}
an expression that can therefore be satisfied for a metric function $h$
whose homogeneous part is given by the Schwarzschild-(A)dS metric
function and whose non-homogeneity is represented by the terms
proportional to $F_4$ and $\tilde{G}_{4,X}$, and where $X$ is
defined implicitly by (\ref{Jmaster}). Unifying all these results,
we conclude that the subclass of actions (\ref{actionbH})
parameterized in terms of $\tilde{G}_2$ and $\tilde{G}_4$ with
\begin{equation}
F_4(X)=\frac{-2 X \tilde{G}_{4, X}(X)+\tilde{G}_4(X)}{4 X^2}, \label{F4expression} 
\end{equation}
that is,
\begin{align}
\begin{split}\label{superaction}
S_{\left\{\tilde{G}_4(X)\right\}}[g, \phi]=\int d^d x \sqrt{-g}\left[a_0+a_1R+\tilde{G}_2(X)+\tilde{G}_4(X)R+\frac{\tilde{G}_{4}(X)}{2\,X}\left((\square \phi)^2-\phi_{\mu \nu} \phi^{\mu \nu}\right)\right. \\
\left.+\left(\frac{-2 X \tilde{G}_{4, X}(X)+\tilde{G}_4(X)}{2 X^2}\right)\left(  \square \phi\, \phi^\mu \phi_{\mu \nu} \phi^\nu -  \phi_\mu \phi^{\mu \nu} \phi_{\nu \rho} \phi^\rho\right)\right],
\end{split}
\end{align}
will admit hairy black hole solutions with primary hair for the
ansatz (\ref{ansatz}), with $f=h$, for a scalar field of the form
\begin{eqnarray}
\phi(t,r)=q\,t\pm \int \sqrt{\frac{q^2}{f(r)^2}-\frac{2
X(r)}{f(r)}}\,dr.
\end{eqnarray}
The metric solution $f=h$ will be the superposition of the
Schwarzschild-(A)dS metric function and a non-homogeneous part, mostly controlled by the primary hair parameter, $q$,
\begin{align}
f(r)=\frac{a_0
r^2}{a_1(d-1)(d-2)}+\kappa-\frac{2M}{r^{d-3}}+\frac{1}{a_1
r^{d-3}}\int \mathcal{H} r^{d-4} dr, \label{metricgen}
\end{align}
where
\begin{equation}
\mathcal{H} =2  q^2  r F_4
X'-\tilde{G}_{4,X}(d-3)\left(\kappa-\frac{q^2}{2X}\right)\left(\frac{\tilde{G}_2}{\tilde{G}_{2,X}}-\frac{\tilde{G}_4}{\tilde{G}_{4,X}}\right).
\label{homocon}
\end{equation}
Notice that since $X \propto q^2$, the absence of the hair in this
expression will be consistent only for $\tilde{G}_2\propto\tilde{G}_4\propto
\sqrt{X}$ (so that $F_4$ and the last bracket in \eqref{homocon} vanish), reducing the solution to the black hole stealth already found in \cite{Bakopoulos:2023fmv}\footnote{The other possibility will be to chose $\tilde{G}_4=$ cst but this is in
contradiction with our construction, see equation (\ref{conditions}).}.

It is also interesting to note that the standard fall-off
$M/r^{d-3}$ can be understood via the Kerr-Schild approach developed
in \cite{Babichev:2020qpr}. For the sake of clarity and compactness,
here we reproduce briefly the arguments as originally presented for
the general case \cite{Babichev:2020qpr}. We start with a seed
configuration of the form
\begin{equation}
ds_0^2=-h_0(r)dt^2+\frac{dr^2}{f_0(r)}+r^2d\Omega_{d-2,\kappa}^2,
\qquad X=X_0(r),
\end{equation}
a solution of the field equations (\ref{eqJ}), (\ref{eqrr}) and
(\ref{eqtt}), where $h_0$ and $f_0$ are mass-independent functions.
Operating with a Kerr-Schild transformation $ds^2=ds_0^2+M
a(r)\,l\otimes l$, where the null geodesic vector field is
$l=dt-dr/(\sqrt{h_0 f_0})$, and requiring invariance of the standard
kinetic term under the transformation, proves to be equivalent to
mapping the original seed functions according to $h_0(r)\to
h(r)=h_0(r)-M a(r)$ and $f_0(r)\to f(r)= f_0(r)(h_0(r)-M
a(r))/h_0(r)$. It is then easy to see that, since $X$ is invariant,
these transformations will map the equations (\ref{eqJ}) and
(\ref{eqrr}) to
\begin{eqnarray*}
\frac{\mathcal{J}^r}{h^2}\to \frac{\mathcal{J}^r}{h^2}+M
(d-2)\frac{f_0}{h_0}\left[r
a^{\prime}+a(d-3)\right]\mathcal{Z}_X,\quad
\frac{\epsilon_{rr}}{h^3}\to \frac{\epsilon_{rr}}{h^3}+M
(d-2)\frac{f_0}{h_0}\left[r a^{\prime}+a(d-3)\right]\mathcal{Z}.
\end{eqnarray*}
Hence, one can conclude that a Kerr-Schild transformation
leaving invariant the kinetic term will be a symmetry of the
independent equations provided the Kerr-Schild function $a(r)$
satisfies the equation $r a^{\prime}(r)+a(r)(d-3)=0$, that is
$a(r)\sim r^{3-d}$.

\subsection{Schwarzschild-like hairy black holes}
A very appealing model that allows for explicit analytic expressions
is the one characterized by $\tilde{G}_2=\lambda\tilde{G}_4$, with $\lambda$ being a constant. As a matter of fact, in this case the explicit
form of the kinetic term is directly identifiable from equation
(\ref{Jmaster}), yielding
\begin{equation}
X(r)=\frac{(d-2)(d-3)q^2}{2\left[\lambda r^2+\kappa
(d-2)(d-3)\right]}. \label{XEH}
\end{equation}
In addition, the non-homogenous contribution of the metric
$\mathcal{H}$ drastically simplifies, and it is simply given by the
beyond Horndeski function $F_4$, which from (\ref{F4expression}) is
shown to be determined in terms of $\tilde{G}_4$ only. In
consequence, the subclass of actions (\ref{superaction})
parameterized in terms of $\tilde{G}_4$, with
$\tilde{G}_2(X)=\lambda\, \tilde{G}_4(X)$, admits a hairy black hole
solution with a scalar field
\begin{eqnarray}
\phi=qt\pm \int \sqrt{\frac{q^2}{f(r)^2}\left(1-\frac{(d-2)(d-3)
f(r)}{\lambda r^2+\kappa (d-2)(d-3)}\right)}\,dr,
\label{scalarfieldsol}
\end{eqnarray}
where the metric reads
\begin{eqnarray}
f(r)=\frac{a_0
r^2}{a_1(d-1)(d-2)}+\kappa-\frac{2M}{r^{d-3}}+\frac{2q^2}{a_1
r^{d-3}}\int r^{d-3}
X'\Bigg[\frac{-2X\tilde{G}_{4,X}+\tilde{G}_4}{4X^2}\Bigg]\, dr.
\label{metric}
\end{eqnarray}
Few comments are in order regarding this hairy black hole solution
as defined by equations (\ref{XEH}-\ref{metric}). Firstly, one can
recognize that the metric solution is a superposition of the
Schwarzschild-(A)dS metric together with a piece proportional to the
scalar hair, $q$. In other words, the scalar hair solution
continuously connects to the Schwarzschild-Tangherlini-(A)dS
solution, providing a scalarized version of the Schwarzschild-(A)dS
black hole in any dimension. Secondly, the integral piece of the
metric solution (\ref{metric}) is, of course, defined modulo an
integration constant, but since this integral is multiplied by a
factor $r^{3-d}$, this "extra" constant can be absorbed into a
redefinition of the mass parameter $M$. Finally, it is desirable
that the metric function behaves asymptotically as the
Schwarzschild-(A)dS metric, that is
\begin{equation}
f(r)\sim \frac{a_0
r^2}{a_1(d-1)(d-2)}+\kappa-\frac{2M}{r^{d-3}}+O\left(\frac{1}{r^{d-3}}\right),
\end{equation}
namely, neither the (A)dS term, nor the mass fall-off are affected by the
inhomogeneous contribution in the metric function. Taking into
consideration (\ref{XEH}) and its derivative, this requirement
translates into the condition
\begin{equation}
\vert \tilde{G}_4-2X\tilde{G}_{4,X}\vert \sim
\frac{1}{r^{\alpha}},\qquad \alpha>2.
\end{equation}
A constraint that particularly affects the use of a $\tilde{G_4}$
function that is linear in $X$, \cite{Bakopoulos:2023fmv}. It is
interesting to remark that to have an everywhere finite
kinetic term (\ref{XEH}), one can simply choose the sign of the
coupling $\lambda$ to be equal to that of the base manifold
curvature, $\mbox{sgn}(\lambda)=\mbox{sgn}(\kappa)$. Moreover, for a
flat base manifold $\kappa=0$, the kinetic term will express a
divergence at the origin $r=0$, however, hidden behind the would-be
event horizon.

\section{Scalarizing theories beyond GR: Einstein-Maxwell, Lovelock and cubic quasitopological}
Having established the scheme behind the scalarization of the
Schwarzschild-(A)dS black hole, we extend our result to other
gravity theories, in particular the cases of Einstein-Maxwell
theory, Lovelock gravity and the so-called cubic quasitopological gravity.
To proceed, we start by complementing the two-parametric
action (\ref{superaction}) with an action depending on the same
metric $g$ and a collection of matter fields, denoted by $\psi_m$,
different from the original beyond Horndeski scalar $\phi$,
yielding\footnote{In the eventual case in which the sector defined
by ${\cal \tilde{L}}_m$ already involves the Einstein-Hilbert piece
(resp. the cosmological constant), we will then consider the action
(\ref{superaction}) with $a_1=0$ (resp. with $a_0=0$) in order
to avoid a repetition of these terms.}
\begin{align}
\begin{split}\label{actionbH2}
S[g, \phi, \psi_m]&=\int d^dx\sqrt{-g}\Bigg[a_0+ a_1 R+
\tilde{G}_2(X)+\tilde{G}_4(X)R+\frac{\tilde{G}_{4}(X)}{2\,X}\Big(\left(\Box\phi\right)^2
-\phi_{\mu\nu}\phi^{\mu\nu}\Big) \\
&+\left(\frac{-2X\tilde{G}_{4,X}(X)+\tilde{G}_4(X)}{2X^2}\right)
\left(  \square \phi\, \phi^\mu \phi_{\mu \nu} \phi^\nu -  \phi_\mu \phi^{\mu \nu} \phi_{\nu \rho} \phi^\rho\right)\Bigg]+ \int
d^dx\sqrt{-g}\, {\cal \tilde{L}}_m(g,\psi_m).
\end{split}
\end{align}
Denoting the field equations coming from the variation of ${\cal
\tilde{L}}_m$ with the metric as $\tilde{\epsilon}_{\mu\nu}$, we
consider the following hypotheses:
\begin{enumerate}[label=(\roman*)]
\item The field equations of the Lagrangian ${\cal \tilde{L}}_m$ admit a
homogeneous static metric solution with purely radial fields of the
form
\begin{eqnarray}
ds^2=-\tilde{f}(r)dt^2+\frac{dr^2}{\tilde{f}(r)}+r^2d\Omega_{d-2,\kappa}^2,\qquad
\psi_m=\psi_m(r). \label{homm}
\end{eqnarray}
\item The field equations $\tilde{\epsilon}_{tt}$ and
$\tilde{\epsilon}_{rr}$ are proportional modulo the field equations
associated to the equations defining the other fields, $\psi_m(r)$.
\end{enumerate}
It is easy now to prove that the full action (\ref{actionbH2}) will
admit a hairy solution of the form
\begin{align}
\begin{split}\label{solgen}
ds^2&=-f(r)dt^2+\frac{dr^2}{f(r)}+r^2d\Omega_{d-2,\kappa}^2,\quad
\hspace{0.1cm} \phi(t,r)=q\,t\pm \int
\sqrt{\frac{q^2}{f(r)^2}-\frac{2 X(r)}{f(r)}}\,dr,\\
\psi_m&=\psi_m(r),
\end{split}
\end{align}
with $X$ defined implicitly by (\ref{Jmaster}), and where the
metric function $f$ will satisfy the following non-homogeneous
differential equation, {\small
\begin{align}
\begin{split}\label{comerr22}
&\tilde{\epsilon}_{rr}(r, f, f',f'', \cdots, \psi_m,
\psi_m',\psi_m''\cdots ) -a_0
r^2+a_1(d-2)\Bigg[r f'+(d-3)f-\kappa (d-3)\Bigg]= \\
&2 (d-2) q^2 r F_4 X'
-G_{4,X}(d-2)(d-3)\left(\kappa-\frac{q^2}{2X}\right)\Big(\frac{\tilde{G}_2}{\tilde{G}_{2,X}}-\frac{\tilde{G}_4}{\tilde{G}_{4,X}}\Big).
\end{split}
\end{align}}
In what follows, we will provide three
representative examples in addition to a simple counterexample
that allows for a deeper understanding of the hypotheses.

\subsection{Black holes with primary hair in Einstein-Maxwell theory}
The simplest case for a theory of the form  ${\cal \tilde{L}}_m$
satisfying the hypotheses $(i)$ and $(ii)$ is the one of
Einstein-Maxwell theory. Indeed, as we know, there exists a simple
spherically symmetric solution, the Reissner-Nordström black hole,
which is actually found from a set of field equations satisfying
that $\tilde{\epsilon}_{tt}\sim\tilde{\epsilon}_{rr}$. Solving
(\ref{comerr22}) and the corresponding Maxwell equations we obtain
\begin{align}
\begin{split}
f(r)&=\frac{a_0
r^2}{a_1(d-1)(d-2)}+\kappa-\frac{2M}{r^{d-3}}+\frac{2Q^2}{(d-2)(d-3)r^{2(d-3)}}\\
&+\frac{1}{a_1 r^{d-3}}\int r^{d-4}\Bigg[2  q^2  r F_4 X'
-\tilde{G}_{4,X}(d-3)\left(\kappa-\frac{q^2}{2X}\right)\Big(\frac{\tilde{G}_2}{\tilde{G}_{2,X}}-\frac{\tilde{G}_4}{\tilde{G}_{4,X}}\Big)
\Bigg]\, dr,
\end{split}\\
A_0(r)&=\frac{Q}{(d-3)r^{d-3}},
\end{align}
where, again, it is evident how the primary hair is added on top of
the bald initial solution. It is interesting to remark that, in this
case, as well as in the subsequent cases, stealth black holes are
simply found by considering $\tilde{G}_2\propto
\tilde{G}_4\propto\sqrt{X}$. As noticed in (\ref{homocon}) for such
a choice of the theory functions, the non-homogeneous source always
vanishes. In this particular subsection, this black hole corresponds
to a charged stealth solution defined on top of the
Reissner-Nordström metric.

\subsection{Black holes with primary hair in Lovelock gravity}
Another appealing example in which the hypothesis $(i)$ and $(ii)$
are fulfilled, is the one of Lovelock gravity \cite{Lovelock:1971yv}. Lovelock theory
represents the natural higher dimensional generalization of Einstein's
theory; therefore it is to be expected that condition $(ii)$ will indeed
hold. In addition, it is known that in arbitrary dimension, and for
the complete series representing the whole tower of curvature
invariants, up to order $[(d-2)/2]$, a spherically symmetric solution
always exists, at least implicitly given by the so-called Wheeler
polynomial, of which its most representative explicit case is given
by the Boulware-Deser black hole \cite{Boulware:1985wk}, the
spherically symmetric solution of the Einstein-Gauss-Bonnet system.

In consequence, considering the Lagrangian of Lovelock gravity of
order $k$ (in which the zero and first order terms represent the
cosmological constant and Einstein-Hilbert contributions),
\begin{equation}
{\cal
\tilde{L}}_m(g)=\sum_{k=0}^{[\frac{d-1}{2}]}a_k\frac{(2k)!}{2^k}\delta^{\mu_1}_{[\alpha_1}
\delta^{\nu_1}_{\beta_1}\cdots\delta^{\mu_k}_{\alpha_k}
\delta^{\nu_k}_{\beta_k]} \prod_{r=1}^{k}
R^{\alpha_r\beta_r}_{\qquad \mu_r\nu_r},
\end{equation}
the metric function describing the hairy generalization of the
general Lovelock black hole is implicitly given by a root of the
following generalized Wheeler polynomial
\begin{align}
\begin{split} \label{wheelermod}
\sum_{k=0}^{[\frac{d-1}{2}]}\frac{a_k
(d-1)!}{(d-2k-1)!}\left(\frac{\kappa-f(r)}{r^2}\right)^k&=\frac{2M}{r^{d-1}}-\frac{1}{r^{d-1}}\int r^{d-4}\Bigg[2  q^2  r F_4 X' \\
&-\tilde{G}_{4,X}(d-3)\left(\kappa-\frac{q^2}{2X}\right)\Big(\frac{\tilde{G}_2}{\tilde{G}_{2,X}}-\frac{\tilde{G}_4}{\tilde{G}_{4,X}}\Big)
\Bigg]\, dr.
\end{split}
\end{align}
From the polynomial (\ref{wheelermod}), explicit cases are easily
obtainable, up to the solution of the corresponding algebraic equation. Therefore, solutions like the Boulware-Deser black hole with primary hair or hairy generalizations with even higher corrections in the curvature, such as the cases in which a degenerate vacuum
arises, can be straightforwardly studied. In addition, charged solutions follow with ease.

\subsection{Black holes with primary hair in cubic quasitopological Gravity}
Let us now shift our focus to an alternative yet intriguing higher curvature order gravity, namely, cubic quasitopological gravity. These theories, originally constructed in \cite{Oliva:2010eb}, represent a class of higher curvature order gravity theories that, when assuming spherical symmetry, result in second-order field equations. Notably, they deviate from the general case of Lovelock theories, particularly in any odd dimension. Extensively investigated in the literature \cite{Cisterna:2017umf,Bueno:2016xff,Bueno:2016lrh,Bueno:2016ypa,Hennigar:2017ego,Cisterna:2018tgx,Moreno:2023rfl,Bueno:2019ltp}, quasitopological gravities provide a compelling framework for our exploration.
In this context, we embark on a complementary approach by augmenting the cubic quasitopological Lagrangian with action (\ref{actionbH2}) for the case of $d=5$. This extension aims to introduce primary hair to the already identified black holes within this quasitopological model \cite{Oliva:2010eb}
\begin{align}
\begin{split}
{\cal \tilde{L}}_m(g)&=a_2{\cal G}+a_3\left(-\frac{7}{6}R_{\ \
\lambda \rho}^{\mu \nu}R_{\ \ \nu \tau}^{\lambda \sigma}R_{\ \ \mu
\sigma}^{\rho \tau} - R_{\mu \nu}^{\ \ \lambda \rho}
R_{\lambda \rho}^{\ \ \nu \sigma}R^{\mu}_{\ \sigma} - \frac{1}{2}R_{\mu \nu}^{\ \ \lambda \rho}R_{\ \lambda}^{\mu}R_{\ \rho}^{\nu} \right. \\
& \left. + \frac{1}{3}R_{\ \nu}^{\mu}R_{\ \lambda}^{\nu}R_{\
\mu}^{\lambda} - \frac{1}{2}RR_{\ \nu}^{\mu}R_{\ \mu}^{\nu} +
\frac{1}{12}R^{3}\right),
\end{split}
\end{align}
where ${\cal G}=R^2-4 R_{\mu \nu} R^{\mu \nu}+R_{\mu \nu \lambda
\rho} R^{\mu \nu \lambda \rho}$ is the Gauss-Bonnet density. In \cite{Oliva:2010eb}, this cubic theory was shown to have second order traced field equations, and admit black hole solutions fitting our hypotheses. It is then easy to see that for
the coupled system, the metric function solution, $f$, will satisfy
the following non-homogeneous cubic equation
\begin{align}
\begin{split}
&\frac{a_3}{3} \frac{\left[\kappa-f(r)\right]^{3}}{r^{2}}+a_2\left[2f(r)\,\left(2 \kappa-f(r)\right)\right]+a_1\,r^{2} \left(f(r)-\kappa\right)-\frac{a_0 r^4}{12}\\
&= - 2M + \int r\Bigg[2  q^2  r F_4 X'
-\tilde{G}_{4,X}(d-3)\left(\kappa-\frac{q^2}{2X}\right)\Big(\frac{\tilde{G}_2}{\tilde{G}_{2,X}}-\frac{\tilde{G}_4}{\tilde{G}_{4,X}}\Big)
\Bigg]\, dr
\end{split}
\end{align}
and, hence converting the black hole solution of
\cite{Oliva:2010eb} to a hairy solution of the full theory. A
similar construction can also be achieved in any odd dimension
$d=2p-1$, by considering the general Lagrangian \cite{Oliva:2010eb}
\begin{equation}
\mathcal{L}_{p} = \frac{1}{2^p} \left( \frac{1}{d-2p+1} \right)
\delta^{\mu_1 \nu_1 \dots \mu_p \nu_p}_{\lambda_1 \rho_1 \dots
\lambda_p \rho_p} \left( C^{\lambda_1 \rho_1}_{\mu_1 \nu_1} \dots
C^{\lambda_p \rho_p}_{\mu_p \nu_p} - R^{\lambda_1 \rho_1}_{\mu_1
\nu_1} \dots R^{\lambda_p \rho_p}_{\mu_p \nu_p} \right) - \gamma_p
C^{\mu_1 \nu_1}_{\mu_p \nu_p}C^{\mu_1 \nu_1}_{\mu_2 \nu_2} \dots
C^{\mu_{p-1} \nu_{p-1}}_{\mu_p \nu_p},
\end{equation}
where
\begin{equation}
\gamma_p = \frac{(d-4)!}{(d-2p+1)!} \frac{\left[
p(p-2)D(d-3)+p(p+1)(d-3)+(d-2p)(d-2p-1) \right]}{\left[ (d-3)^{p-1}
(d-2)^{p-1} + 2^{p-1} - 2(3-D)^{p-1} \right]},
\end{equation}
and where $C_{\mu \nu \lambda \rho}$ denotes the Weyl tensor. As before, we set $a_0$ and
$a_1$ to zero in \eqref{actionbH2}. Then, the metric solution $f(r)$
satisfies the following polynomial equation, similar to the pure
Lovelock case:
\begin{align}
\begin{split}
&\sum_{k=0}^{p}(-1)^k\binom{p}{k}\overline{a}_k\left(\frac{\kappa-f(r)}{r^2}\right)^k \\
&=\frac{2M}{r^{d-1}}-\frac{1}{r^{d-1}}\int r^{d-4}\Bigg[2  q^2  r
F_4 X'
-\tilde{G}_{4,X}(d-3)\left(\kappa-\frac{q^2}{2X}\right)\Big(\frac{\tilde{G}_2}{\tilde{G}_{2,X}}-\frac{\tilde{G}_4}{\tilde{G}_{4,X}}\Big)
\Bigg]\, dr.
\end{split}
\end{align}
Note that the dimensional factor of the term coming from
$\mathcal{L}_{p}$ has a slightly different form than those of the
Lovelock terms. As we have already mentioned, in odd dimensions, these theories do not coincide. Hence, we have absorbed all factors into their
corresponding coupling constants and rescaled them to obtain a more
convenient form. This form allows for another simplification when
the rescaled coupling constants are such that
\begin{align}
\overline{a}_0 =
\frac{\overline{a}_{p-1}^p}{\overline{a}_p^{p-1}},\qquad
\overline{a}_1 =
\frac{\overline{a}_{p-1}^{p-1}}{\overline{a}_p^{p-2}},\qquad
\overline{a}_2 =
\frac{\overline{a}_{p-1}^{p-2}}{\overline{a}_p^{p-3}},\cdots,
\overline{a}_{p-2} = \frac{\overline{a}_{p-1}^2}{\overline{a}_p}.
\end{align}
In this case, the polynomial equation of order $p$ reads
\begin{align}
\begin{split}
&\frac{\left\{ r^2 \overline{a}_{p-1} - \overline{a}_p [\kappa - f(r)] \right\}^p}{r^{2p}\overline{\alpha}_p^{p-1}} \\
&=\frac{2M}{r^{d-1}}-\frac{1}{r^{d-1}}\int r^{d-4}\Bigg[2  q^2  r
F_4 X'
-\tilde{G}_{4,X}(d-3)\left(\kappa-\frac{q^2}{2X}\right)\Big(\frac{\tilde{G}_2}{\tilde{G}_{2,X}}-\frac{\tilde{G}_4}{\tilde{G}_{4,X}}\Big)
\Bigg]\, dr.
\end{split}
\end{align}

\subsection{A counterexample with the vacuum conformal gravity
solution}
To appreciate our working hypotheses, we shall provide a
counterexample, the case of Conformal Gravity \cite{Mannheim:1988dj}. There exists an elementary example of a static
solution in the homogeneous form (\ref{homm}) for which the field
equations $\tilde{\epsilon}_{tt}$ and $\tilde{\epsilon}_{tt}$ are
not proportional. Consider the action (\ref{actionbH2}) in $d=4$
with $a_0=a_1=0$ and
$$
{\cal \tilde{L}}_m(g)= C_{\lambda \mu \nu \kappa} C^{\lambda \mu \nu
\kappa},
$$
where, as before, $C_{\lambda \mu \nu \kappa}$ is the Weyl tensor. Indeed, one
can see that in this case, the hypothesis (ii) is broken since
$$
\frac{\tilde{\epsilon}_{t t}}{f^2(r)}+\tilde{\epsilon}_{r
r}=-\frac{8}{3} \frac{f^{\prime \prime \prime }(r)}{r}-\frac{2}{3}
f^{\prime \prime \prime \prime}(r).
$$
Note that, in the vacuum case, the metric solution
\cite{Mannheim:1988dj} is such that the right-hand side vanishes
identically, and hence there is no consequence of
$\tilde{\epsilon}_{t t}$ not being proportional to
$\tilde{\epsilon}_{t t}$. However, in general, in our
construction, the non-homogeneity acquired from the primary hair will not be such that the right-hand side of the previous equation vanishes.

\section{Further comments}
In this work, we have explored a scalar-tensor theory within the so-called  beyond Horndeski gravity, a theory that accommodates higher-order terms while preserving the propagation of healthy degrees of freedom \cite{Gleyzes:2014dya}. We have deliberately confined our investigation to a scalar theory invariant under a constant shift of the scalar field. In this particular scenario, the presence of a conserved Noether current proves to be an important tool in the search for analytical solutions.
Remarkably, the action (\ref{actionbH}) has demonstrated noteworthy potential in the construction of exact black hole solutions, as evidenced by the recent work of \cite{Bakopoulos:2023fmv}. The solutions presented in this reference have been found for specific coupling functions $G_2$, $G_4$, and $F_4$, and correspond to black holes characterized by primary scalar hair. Notably, these scalarized black holes maintain a continuous connection with the Schwarzschild solution. 

In this investigation, we have extended the framework of \cite{Bakopoulos:2023fmv} in two key directions. Firstly, we considered its arbitrary dimensional extension, and secondly, we explored the circumstances under which the beyond Horndeski model, as considered herein, can be coupled to gravity theories beyond GR to yield hairy black holes with primary hair. Concerning the former, we have identified the most comprehensive class of actions (\ref{actionbH}) that facilitates the emergence of black holes with primary scalar hair that are continuously connecting to the Schwarzschild-Tangherlini-(A)dS solution.
In relation to the latter point, we have precisely identified the possible alternative gravitational theories (with possibly extra dynamical fields different from the scalar field) that can be coupled with our beyond Horndeski model to generate scalarized black holes. Subsequently, we have demonstrated how the Reissner-Nordström black hole of Einstein-Maxwell, as well as the static black hole configurations within general Lovelock and cubic quasitopological gravities, can be promoted to configurations featuring primary hair.
As in \cite{Bakopoulos:2023fmv}, the hair of our solutions is given by the integration constant $q$. This constant, which accompanies the linear time dependence of the scalar field, ensures the regularity of the scalar field throughout the spacetime. It is essential to note that, although all our solutions exhibit a continuous limit with their bald counterparts, they do not necessarily represent a linear superposition.

In the broader context of scalar-tensor theories, such as Horndeski, beyond Horndeski, or DHOST theories \cite{Langlois:2015cwa, Langlois:2015skt, Crisostomi:2016czh}, it is known  that certain specific sectors exhibit black hole solutions that are distinctly disconnected from the Schwarzschild solution. In light of this observation, it would be appealing to categorize all these scalar-tensor theories according to whether or not they support solutions that are continuously connecting to vacuum solutions of General Relativity.
To pursue this task, insights coming from the Kerr-Schild framework developed in \cite{Babichev:2020qpr} could be useful. In this reference, shift-invariant DHOST theories invariant under a Kerr-Schild transformation that preserves the invariance of the kinetic term were identified (with a scalar field possibly linear in the time coordinate). Particularly, those theories featuring a standard fall-off mass term in their static black hole solutions emerge as potential candidates for accommodating hairy black holes that are continuously connected with the Schwarzschild solution.

Finally, the availability of exact black hole spacetimes with primary hair opens avenues for several analyses that can be done analytically. Notably, it becomes pertinent to conduct in-depth investigations into the thermodynamic properties of these spacetimes, compute the corresponding black hole charges, establish methodologies for regularizing a given theory action, and substantiate the validity of the first law. Then several applications of these solutions can be performed along the lines of the AdS/CFT conjecture or black hole chemistry \cite{Kubiznak:2016qmn}. Additionally, a comprehensive analysis of the causal structure, geodesics, and algebraic classification promises a more nuanced understanding of the interplay between vacuum and hairy black holes with primary hair. We anticipate presenting findings on these aspects in the near future.

\section*{Acknowledgments}

We express our gratitude to Eloy Ay\'on-Beato, Christos Charmousis, Luis Guajardo, Nicolas Lecoeur and Julio
Oliva for interesting discussions regarding the topics exposed here.
The work of A.C. is partially supported by Fondecyt Grant 1210500
and Primus grant PRIMUS/23/SCI/005 from Charles University. The work
of M.H. is partially funded by Fondecyt Grant 1210889. The work of
U.H. is partially supported by by ANID grant 21231297.
A.C. would like to thank the Paris Aéroport Roissy Charles de Gaulle for providing the necessary environment to finish this work. A.C. and M.H. would like also to thank l'Institut Pascal of University Paris-Saclay where this work has been initiated. 

\section*{Appendix: Concrete examples}
For the sake of concreteness, we will give some explicit cases. We start by considering \eqref{superaction}
\begin{align}
    \tilde{G}_2(X) =& -\frac{\lambda(d-3)(d-2)}{2 n-1} X^n,\\
    \tilde{G}_4(X) =& - \frac{\lambda}{2n - 1} X^n,\\
    F_4(X) =& \frac{1}{4} \lambda X^{n-2},
\end{align}
where evidently $\tilde{G}_2\propto \tilde{G}_4$. Recall that case $n=\frac{1}{2}$
deserves a particular attention as already noticed in
\cite{Bakopoulos:2023fmv} (see below).
Further, for convenience, we have fixed the
constants $a_0 = 0$ and $a_1=1$. This provides a kinetic term $X$ of the form
\begin{equation}
    X = \frac{q^2}{2 \left(\kappa+r^2\right)},
\end{equation}
and the solution of the metric function for the case of GR then reads
\begin{align}
    f(r) &= \kappa - \frac{2M}{r^{d-3}} - \frac{2 \lambda}{r^{d-3}}\int r^{d-2} X^n \, dr = \kappa - \frac{2M}{r^{d-3}} + \frac{\lambda q^2}{2 (n-1)} \left[X^{n-1} - \frac{(d-3)}{r^{d-3}}\int r^{d-4} X^{n-1} \, dr \right] \\
    &=\kappa - \frac{2M}{r^{d-3}} - \frac{2^{1-n} \lambda \kappa^{-n} q^{2 n}}{d-1} r^2 \, _2F_1\left(\frac{d-1}{2},n;\frac{d+1}{2};-\frac{r^2}{\kappa}\right),
\end{align}
where $\, _2F_1$ is the Gaussian hypergeometric function. In
particular, for $n$ being an integer, this takes an even more simple
form, for example, for $n=3$ and in four dimensions we obtain
\begin{equation}
    f(r) = \kappa - \frac{2M}{r} - \frac{\lambda q^6 }{32 \kappa}\left[ \frac{(r^2 - \kappa)}{(r^2 + \kappa)^2} + \frac{1}{r \kappa^{1/2}}\arctan{\left( \frac{r}{\sqrt{\kappa}} \right)} \right].
\end{equation}
We can even extend this to include the Gauss-Bonnet term in the
action in the same line as we did in the general Lovelock case. This
yields a general form of the metric function,
\begin{small}
\begin{equation}
    f = \kappa+\frac{r^2}{2\alpha (d-4) (d-3)} \left(1 \pm \sqrt{1 + \frac{8\alpha M (d-4) (d-3)}{r^{d-1}} + \frac{8 \lambda\alpha (d-4) (d-3)}{r^{d-1}} \int r^{d-2} X^n \, dr}\right),
\end{equation}
\end{small}
where $\alpha$ is the coupling constant of the Gauss-Bonnet term.
Carefully taking the limit $\alpha \rightarrow 0$ (in the case of
the lower sign in front of the square root), this reduces to the
Schwarzschild-like case from before. For example, let us give the
specific form of the metric function in five dimensions for $n=3$:
\begin{equation}
    f = \kappa+\frac{r^2}{4\alpha} \left(1 \pm \sqrt{1 + \frac{16\alpha}{r^{4}}\left[M - \lambda \frac{q^6 \left(\kappa+2 r^2\right)}{32 \left(\kappa+r^2\right)^2}\right]}\right).
\end{equation}
To conclude, as already mentioned for $\tilde{G}_4(X) \propto \sqrt{X}$, the
coupling function $F_4$, as defined by (\ref{F4expression}), vanishes
identically, and consequently the non-homogeneity disappears. Hence,
in the case of Einstein-Gauss-Bonnet gravity and for
\begin{equation}
\tilde{G}_2(X) =-\lambda(d-3)(d-2) \sqrt{X},\qquad \tilde{G}_4(X) = -\lambda \sqrt{X},\\
\end{equation}
we end-up with a stealth defined on the Boulware-Deser black hole
solution similar to the one found in \cite{Babichev:2022awg}
\begin{equation}
    f(r) = \kappa + \frac{r^2}{2 \alpha (d-3)(d-4)} \left( 1 \pm \sqrt{ 1+\frac{8 \alpha M (d-3)(d-4)}{r^{d-1}} } \right).
\end{equation}

\end{document}